\documentclass[print,trackchanges]{ati}	
\usepackage{graphicx,times}
\usepackage{lmodern}
\usepackage[sort&compress, numbers,super]{natbib} 
\bibpunct[,]{[}{]}{,}{n}{}{,}			
\usepackage{url}
\usepackage[colorlinks=true,breaklinks=true, linkcolor=red,urlcolor=magenta,citecolor=blue,anchorcolor=green]{hyperref}
\usepackage{caption}
\usepackage{rotating}
\usepackage{CJKutf8}	
\usepackage{fancyhdr}
\renewcommand{\citet}[1]{\citeauthor{#1}(\citeyear{#1})\cite{#1}}	

\usepackage{amsmath}
\usepackage{multirow}    
\usepackage{xcolor} 

\begin{document}

   \title{
   Inversion of CHASE H$\alpha$ Spectral Line during Solar Flares Based on RADYN Dataset via Deep Learning
}


   \author {W. Xu
      \inst{1}
      \ORCID{0000-0002-2919-6989}		
   \and Q. Hao\correspondingAuthor{}
     \inst{1,2}
     \ORCID{0000-0002-9264-6698}
   \and Z. Zheng
     \inst{1,2}
     \ORCID{0009-0006-5180-8052}
   \and J. Hong  
      \inst{3}
      \ORCID{0000-0002-8002-7785}
   \and J. Hu  
      \inst{1,4}
      \ORCID{0009-0000-5660-0572}
   \and Y. Qiu  
      \inst{5}
      \ORCID{0000-0002-1190-0173}
    \and C. Li  
      \inst{1,2,5}
      \ORCID{0000-0001-7693-4908}
    \and M.D. Ding  
      \inst{1,2}
      \ORCID{0000-0002-4978-4972}
    \and C. Fang  
      \inst{1,2}
   }
\correspondent{Qi Hao}	
\correspondentEmail{haoqi@nju.edu.cn}

\institute{School of Astronomy and Space Science, Nanjing University, Nanjing 210023, China; 
          \and
              Key Laboratory of Modern Astronomy and Astrophysics, Ministry of Education, Nanjing 210023, China;
          \and
              Institute for Solar Physics, Department of Astronomy, Stockholm University, AlbaNova University Centre,106 91 Stockholm, Sweden;
          \and
              Software Institute, Nanjing University, Nanjing 210023, China;
          \and
              Institute of Science and Technology for Deep Space Exploration, Suzhou Campus, Nanjing University, Suzhou 215163, China
   }
   \date{Received:~November 11, 2025;   Accepted:~December 24, 2025;  Published Online:~December 25, 2025; 
   \DOI{ati2025087} }				
   \citeinfo {Xu, W. et al.}\volume{3}\issue{3} \pages{1--11}	
   \StartPage{1} 				
   \MonthIssue{May}		
   \copyrights {2026}     			
   \abstract{
Solar flares represent one of the most intense forms of solar activity. Understanding the evolution of physical parameters in the solar atmosphere during flares is key to studying flare mechanisms and improving prediction capabilities. However, directly measuring quantities such as electron number density, temperature, and plasma velocity remains difficult. Here, we introduce a novel fully connected neural network, trained on synthetic data from the Radiative Hydrodynamics Code (RADYN) simulations, to perform rapid inversion of physical parameters from H$\alpha$ spectral profiles. The spectral data were processed to align with the observational resolution of the CHASE satellite, enabling seamless application of the model to real-world observations. Results demonstrate a high degree of consistency with RADYN simulations, achieving low errors under diverse flare conditions. Furthermore, we applied the developed model to analyze CHASE observations of a class X7.1 solar flare on October 1, 2024. The results reveal reasonable spatial and temporal evolution of key parameters throughout different flare phases. This work demonstrates the potential of deep learning techniques for fast and reliable spectral inversion, providing new tools for solar flare diagnostics based on H$\alpha$ data.
}


   \authorrunning{ASTRONOMICAL TECHNIQUES \& INSTRUMENTS }   
   \titlerunning{Xu, W. et al.: ~Prepare a LaTeX Manuscript for ATI }  
   \maketitle
   \setcounter{page}{\Page}	
%
%
  \keywords{ 
Solar Flare --- Spectral Inversion --- Deep Learning	
}

\section{Introduction}
\label{sec:intro}
Solar flares are one of the most intense forms of solar activity \citep{Shibata2011}. They are characterized by sudden, large-scale energy releases in localized regions of the solar atmosphere \citep{Fletcher2011}. This process causes rapid heating and the emission of a broad spectrum of electromagnetic radiation. The wavelength of this radiation spans the entire electromagnetic spectrum. During solar flares, energy is deposited in the photosphere and chromosphere through mechanisms such as ion beam bombardment, thermal conduction, and magnetic reconnection. These processes induce significant changes in the physical parameters, including electron number density, temperature, and plasma velocity. The evolution of these parameters is closely linked to the onset and development of the flare. For instance, rapid changes in plasma velocity fields can trigger magnetic reconnection, which can then initiate solar flares. Accurately measuring and inverting these physical parameters is therefore essential for understanding the mechanisms behind solar flare eruptions and improving flare prediction accuracy. However, directly measuring these parameters is technically challenging. Instead, their variations are often reflected in the observed spectral features. By analyzing strong spectral lines in the visible, near-infrared, and ultraviolet spectra, we can infer the physical conditions. Spectral lines such as H, Ca II and Mg II reveal that the solar chromosphere, the atmospheric layer between the photosphere and the corona, is highly dynamic, finely structured, and complex \citep{Carlsson2019}.

Complex inversion methods based on physical models often rely on intricate physical assumptions and approximations, resulting in high computational costs and limitations in practical application \citep{Osborne2019,Chappell2022}. In this context, machine learning methods offer a promising alternative, enabling rapid and efficient inversion based on large-scale statistical data, independent of physical models. The application of deep learning techniques in solar flare research is emerging as a powerful approach, driven by advances in computational capabilities \citep{Panos2023,Asensio2023,Huang2024,Cao2025}. \citet{Osborne2019} trained an invertible neural network (INN) based on the results of simulations with the one-dimensional radiative hydrodynamics code RADYN. The INN analyses the input spectral H$\alpha$ and Ca II 8542 lines and outputs physical parameters, such as the temperature and electron number density of the solar atmosphere. In validating the analysis of a class M1.1 solar flare, the expected results were obtained. \citet{Chappell2022} used the results of a large number of 3D solar atmosphere simulations as a training set for a convolutional neural network (CNN) called SunnyNet. SunnyNet computes 3D nonlocal thermodynamic equilibrium radiative transfer in optically thick stellar atmospheres and can predict atomic populations in nonlocal thermodynamic equilibrium with reasonable accuracy. It can also predict the temperature and electron number densities in solar flares with reasonable accuracy. SunnyNet can make reasonable predictions for atomic populations in non-local thermodynamic equilibrium and is nearly 100,000 times faster than conventional codes. \citet{Lee2022} inverted H$\alpha$ and Ca II spectra using deep neural network (DNN) and avoided the gradient vanishing problem of DNNs by using a jump-joining method. The final inversion results were more accurate, and it took only 0.3-0.4 ms for each spectral line, which is 250 times faster than the traditional method. The final inversion results are more accurate, taking only 0.3-0.4 ms for each spectral line, which is 250 times faster than the traditional analysis.

The Chinese H$\alpha$ Solar Explorer (CHASE) is China's first solar exploration satellite. It is equipped with a solar space telescope that can make continuous 24-hour observations in a sun-synchronous orbit at an altitude of 517 kilometers \citep{Li2019,Li2022,Li2023}. The successful launch of CHASE has provided a wealth of high-resolution H$\alpha$ spectral data, enabling more detailed studies of solar flares. The H$\alpha$ line during flares provides crucial insights into the underlying physical processes. In this study,  we apply deep learning techniques innovatively to solar flare research. We aim to construct an appropriate fully connected neural network model that can swiftly and accurately invert solar chromospheric parameters, including electron number density, temperature, and plasma velocity, using H$\alpha$ spectral data. We further validate the practical application of our model by analyzing real H$\alpha$ spectral data obtained by the CHASE satellite during solar flares. Our work is expected to lay a solid scientific foundation for studying the physical mechanisms of solar flares and improving the accuracy of flare eruption predictions.

\begin{figure}[ht!]
    \centering
    \includegraphics[width=\linewidth]{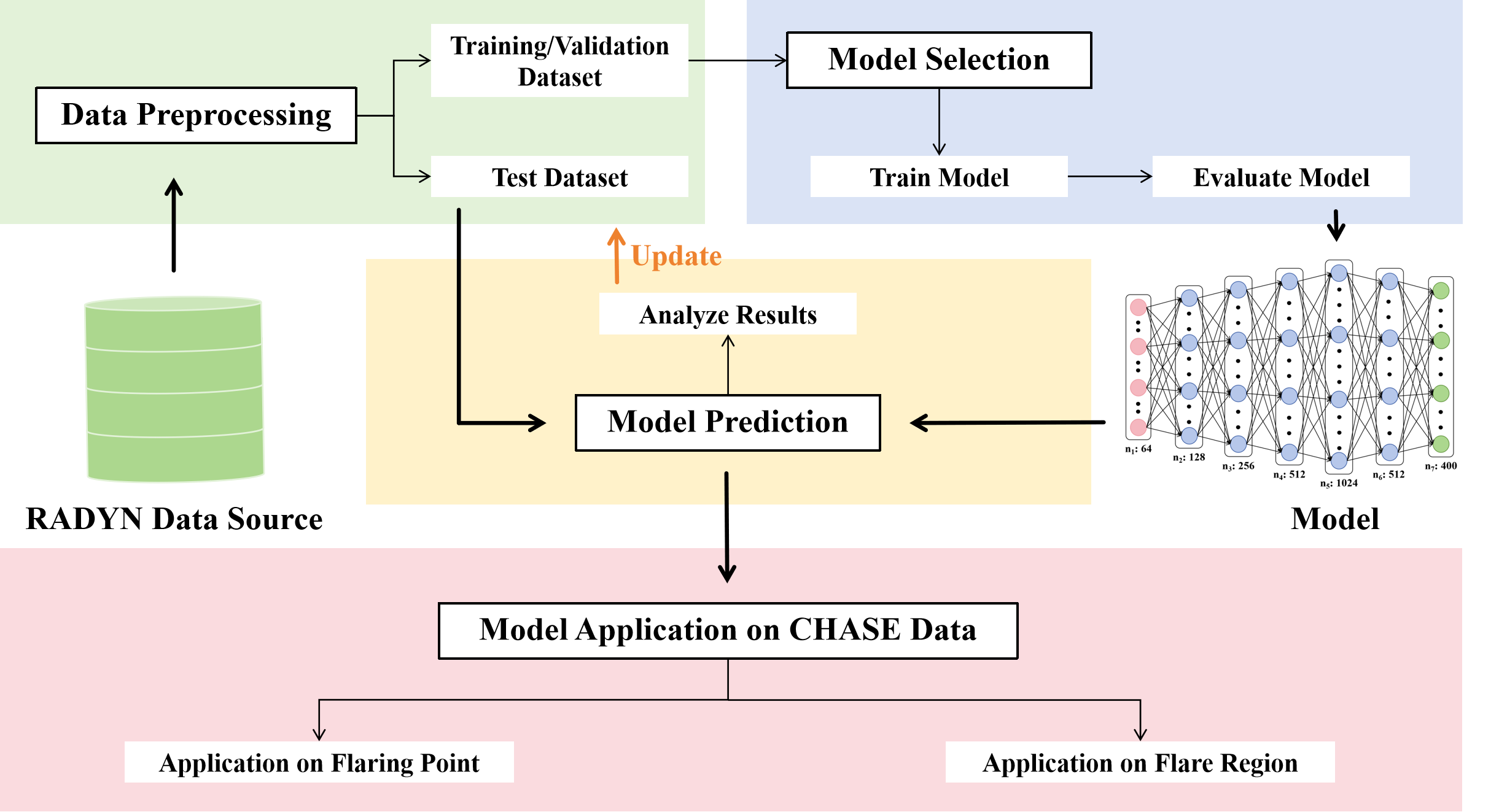}
    \caption{Schematic diagram of the workflow.}
    \label{workflow}
\end{figure}

Figure \ref{workflow} shows the schematic of the workflow. The RADYN data source and data preprocessing are described in Section \ref{sec:dataset}, while the model selection and prediction are described in Section \ref{sec:model}. Section \ref{sec:application} presents the model application on CHASE data. Finally, discussion and conclusion are given in Section \ref{sec:summary}.

\section{Dataset}
\label{sec:dataset}

\subsection{RADYN Flare Dataset}

RADYN is a radiative hydrodynamics code originally developed by \citet{Carlsson1992,Carlsson1997} to simulate acoustic shocks in the chromosphere. It has since been adapted to simulate flare responses in the chromosphere \citep{Allred2005,Allred2015}. RADYN uses an adaptive grid method to solve the radiative hydrodynamics equations for one-dimensional plane-parallel atmospheres. It optimizes data resolution by taking denser samples in regions with larger parameter gradients. This approach ensures that maximum information is retained within the same storage capacity, making it particularly effective for modeling the complex dynamics of solar flares.

The dataset used in our study was from the RADYN F-CHROMA database\footnote{\url{https://star.pst.qub.ac.uk/wiki/public/solarmodels/start.html}}\citep{Carlsson2023}. The dataset comprises 79 flare models, characterized by different spectral indices, cutoff energies, and total energy of the beam electrons that are injected at the flare loop top as an energy source. The energy flux of the beam electrons has a triangular shape as a function of time, with 10 s increase and 10 s decrease. Thus, a total energy of 1$\times$10$^{12}$ erg cm$^{-2}$ corresponds to a maximum energy flux of 1$\times$10$^{11}$ erg cm$^{-2}$ s$^{-1}$.
Each flare model spans 50 seconds, and data are recorded at 0.1-second intervals, resulting in 39,574 time slices. Each data entry includes one-dimensional H$\alpha$ spectral data, as well as the corresponding height distributions of physical parameters such as electron number density, temperature, and plasma velocity. This comprehensive dataset is a valuable resource for analyzing solar flare dynamics. It enables us to explore the relationships between spectral variations and the underlying physical processes in the chromosphere. Including multiple flare configurations and detailed, time-resolved data allows us to robustly analyze flare-induced changes in the solar atmosphere. This enhances our understanding of flare mechanisms and their impact on solar activity.

\subsection{Preprocessing of Spectra}

To effectively apply our model to the real spectra obtained by the CHASE satellite, we pre-process the spectral data from the RADYN dataset. This involves applying a point spread function (PSF) convolution to the original spectral data to simulate the resolution of the CHASE spectrograph. The selected wavelength range is 6562.82 Å ± 1.5 Å, which corresponds to 64 wavelength points in the CHASE data. After applying the PSF, the spectral data is interpolated to a coarser grid to match the CHASE observation features.

\subsection{Preprocessing of Physical Parameters}

We perform interpolation on the distribution data of physical parameters, such as electron number density, temperature, and plasma velocity, we perform interpolation to create smooth profiles suitable for neural network training and inversion. Specifically, we select 400 equally spaced points within the height range of 0-2000 km. This process ensures that the physical parameter distributions are well-resolved and compatible with the required spatial resolution required for subsequent model applications.

\begin{figure}[ht!]
    \centering
    \includegraphics[width=\linewidth]{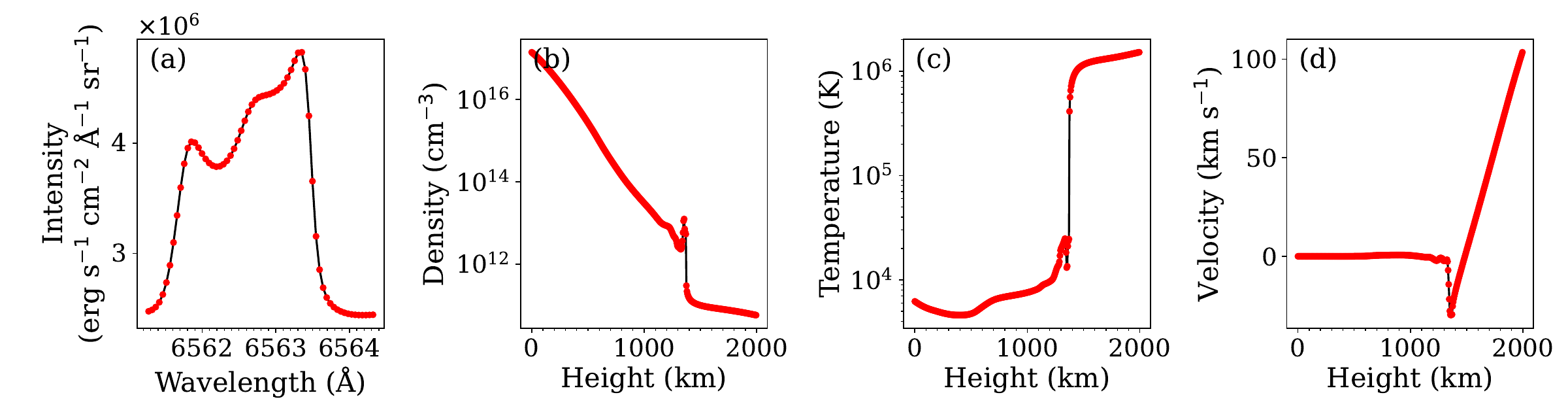}
    \caption{An example of preprocessed RADYN simulation data. (a) Spectrum; (b) Electron number density distribution; (c) Temperature distribution; (d) Plasma velocity distribution. Red dots indicate interpolation points.}
    \label{Distribution}
\end{figure}

Figure \ref{Distribution} shows an example of preprocessed RADYN simulation data derived from the impulsive phase. Figures \ref{Distribution} (a--d) show the preprocessed spectrum, electron number density distribution, temperature distribution, and plasma velocity distribution, respectively.

\section{Deep Learning Spectral Inversion Model}
\label{sec:model}

\subsection{Neural Network Architecture}

We employ a fully connected neural network (FCNN) to learn the nonlinear mapping from H$\alpha$ spectral profiles to the corresponding height-dependent physical parameters. The network architecture consists of an input layer, multiple hidden layers with nonlinear activation functions, and an output layer. Each layer is densely connected, meaning every neuron in one layer is connected to all neurons in the subsequent layer. The architecture is selected to balance model capacity and generalization performance, based on preliminary experiments. The layer widths are empirically chosen to progressively expand and contract the feature space, allowing the network to learn both low-level and high-level representations from the spectral input.

The input to the model is a 64-dimensional vector representing the preprocessed H$\alpha$ spectral intensities, and the output is a 400-dimensional vector corresponding to the vertical distribution of a given physical quantity (i.e., electron number density, temperature, or plasma velocity). Three separate models with the same architecture are trained independently to invert each physical parameter. To enable the network to capture complex spectral–physical relationships while mitigating overfitting, we adopt a moderately deep architecture with PReLU\citep{He2015} activation functions and dropout regularization between layers. In addition, weight decay is applied during training to further constrain model complexity. Table \ref{par} specifically demonstrates the FCNN architecture.

\begin{table}[h!]
  \begin{center}
  \caption{The fully connected neural network architecture.}
    \begin{tabular}{cc} 
      \hline
      Layer/Operation & Shape of Output \\
      \hline
         Input 		& 64 \\
         Linear (64→128) & 128 \\
         PReLU & 128 \\
         Dropout (dropout\_rate=0.2) & 128 \\
         Linear (128→256) & 256 \\
         PReLU & 256 \\
         Dropout (dropout\_rate=0.2) & 256 \\
         Linear (256→512) & 512 \\
         PReLU & 512 \\
         Dropout (dropout\_rate=0.2) & 512 \\
         Linear (512→1024) & 1024 \\
         PReLU & 1024 \\
         Dropout (dropout\_rate=0.2) & 1024 \\
         Linear (1024→512) & 512 \\
         PReLU & 512 \\
         Dropout (dropout\_rate=0.2) & 512 \\
         Linear (512→400) & 400 \\
    \hline
    \end{tabular}\label{par}
  \end{center}
\end{table}
    
\subsection{Training Process}

We use the PyTorch \citep{Paszke2019} deep learning library to build and train models. The models are trained using the AdamW \citep{Loshchilov2017} optimizer, which combines the benefits of momentum and adaptive learning rates, efficiently navigating the high-dimensional parameter space. To optimize model performance, we implement a learning rate decay mechanism to stabilize convergence. We set the initial learning rate to 0.001 and decay it to 0.7 after every 100 iterations. Mean Squared Error (MSE) is used as the loss function. The size of the training batch is set to 128. 

The dataset is divided into a training set, a validation set, and a test set. The RADYN dataset consists of 79 events with four different values of total energy (3$\times 10 ^{10}$ erg $cm^{-2}$, 1$\times 10 ^{11}$ erg $cm^{-2}$, 3$\times 10 ^{11}$ erg $cm^{-2}$, and 1$\times 10 ^{12}$ erg $cm^{-2}$). For each value set, we randomly select four events to form the test set, for a total of sixteen events (about 20\% of the data). This strategy ensures the test events are completely separate for unbiased evaluation. The remaining events are then randomly split 7:1 between the training and validation sets. This results in an  approximate general distribution of 70\% training data (27,613 time slices), 10\% validation data (3945 time slices) and 20\% test data (8016 time slices).

The models undergo 800 iterations on both the training and validation datasets. Throughout this process, training and validation losses steadily decline and ultimately converge, demonstrating effective model fitting to both datasets.

\subsection{Performance Evaluation}

The inversion performance of the models is systematically evaluated using comprehensive error metrics on the training, validation, and test sets. Table \ref{evaluation} presents the quantitative assessment of the electron number density, temperature, and plasma velocity inversion models, respectively, through the root mean square error (RMSE) and symmetric mean absolute percentage error (sMAPE) metrics.

RMSE is a commonly used metric in regression analysis that measures the standard deviation of prediction errors. It is sensitive to large deviations and provides an overall measure of model accuracy.

\begin{equation}
\text{RMSE} = \sqrt{ \frac{1}{n} \sum_{i=1}^{n} \left( \hat{y}_i - y_i \right)^2 },
\end{equation}
where $\hat{y}_i$ and $y_i$ denote the predicted and true values at the i-th grid point, and n is the total number of spatial points.

We also use sMAPE to complement RMSE, which is scale-dependent. sMAPE is defined as follows:

\begin{equation}
\text{sMAPE} = \frac{100\%}{n} \sum_{i=1}^{n} \frac{ \left| \hat{y}_i - y_i \right| }{ \left( |\hat{y}_i| + |y_i| \right) / 2 }.
\end{equation}

sMAPE expresses errors as percentages, providing a scale-independent evaluation that is ideal for comparing variables with different units and magnitudes. Unlike traditional mean absolute percentage error, sMAPE robustly handles near-zero values and treats over- and under-predictions symmetrically, making it ideal for physical parameter inversion. 

Together, RMSE and sMAPE provide a balanced view of both absolute and relative model performance, offering a more comprehensive assessment of the inversion accuracy across different physical quantities. 

For electron number density inversion, the model performs consistently on the training and validation sets and exhibits slightly degraded, yet reasonable, performance on the test set. Similar patterns are observed for temperature and plasma velocity inversions, where test set errors are approximately 1.5 to 2.3 times higher than training/validation errors, reflecting the expected generalization gap.

\begin{figure}[ht!]
    \centering
    \includegraphics[width=\linewidth]{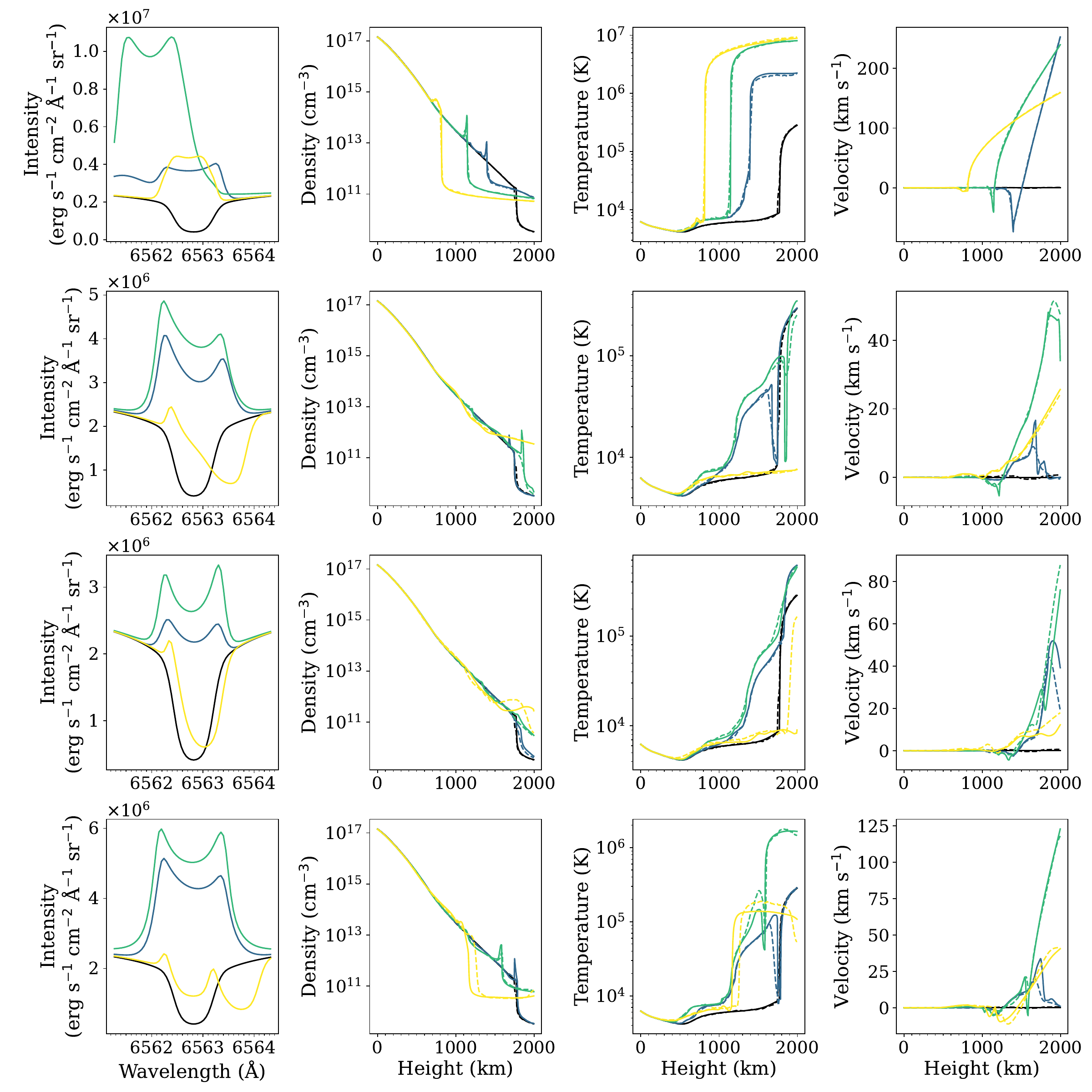}
    \caption{Inversion test results using RADYN simulation data. The four rows from top to bottom correspond to simulated flare events with total energies of 3$\times 10 ^{10}$ erg $cm^{-2}$, 1$\times 10 ^{11}$ erg $cm^{-2}$, 3$\times 10 ^{11}$ erg $cm^{-2}$, and 1$\times 10 ^{12}$ erg $cm^{-2}$, respectively. The four columns from left to right represent spectra, electron number density distributions, temperature distributions, and plasma velocity distributions, respectively. Different colors indicate distinct evolutionary phases: background (black), rising phase (blue), impulsive phase (green), and decay phase (yellow), respectively. Solid lines represent the height distribution of simulated physical parameters, and dashed lines represent the height distribution of the inverted physical parameters.}
    \label{Performance on RADYN Spectra}
\end{figure}

Figure \ref{Performance on RADYN Spectra} provides a detailed comparison of the simulated and inverted physical parameters, showcasing representative test cases from all four total energies (3$\times 10 ^{10}$ erg $cm^{-2}$, 1$\times 10 ^{11}$ erg $cm^{-2}$, 3$\times 10 ^{11}$ erg $cm^{-2}$, and 1$\times 10 ^{12}$ erg $cm^{-2}$) in the test set and further illustrates the inversion performance. Different colors indicate distinct evolutionary phases: background, rising phase, impulsive phase, and decay phase. 

\begin{table*}[]
	\centering
    \caption{RMSE and sMAPE evaluations of the electron number density, temperature, and plasma velocity inversion models.}
	\begin{tabular}{ccccc}
        \hline
	&  & Training Set & Validation Set & Test Set \\
        \hline
	\multirow{2}{*}{Electron Number Density} & RMSE (cm$^{-3}$) & 3.81$\times 10 ^{13}$ & 3.83$\times 10 ^{13}$ & 5.47$\times 10 ^{13}$ \\
		& sMAPE & 1.96\% & 2.02\% & 6.32\% \\
        \hline
	\multirow{2}{*}{Temperature} & RMSE (K) & 2.05$\times 10 ^{4}$ & 2.06$\times 10 ^{4}$ & 3.03$\times 10 ^{4}$ \\
		& sMAPE & 3.97\% & 4.11\% & 11.31\% \\
        \hline
	\multirow{2}{*}{Plasma Velocity} & RMSE (km s$^{-1}$) & 12.67 & 12.60 & 20.96 \\
		& sMAPE & 9.52\% & 9.82\% & 18.47\% \\
        \hline
	\end{tabular}
	\label{evaluation}
\end{table*}

Regarding electron number density distributions, the model shows excellent agreement with the simulated profiles across the atmospheric layers, with minor discrepancies appearing at certain heights. The temperature inversion model performs particularly well, accurately reproducing the absolute values and height-dependent variations across all total energies and flare phases. Although plasma velocity profiles show slightly larger discrepancies compared to the other parameters, consistent with the quantitative metrics in Table \ref{evaluation}, the model nonetheless captures essential dynamical features and reproduces physically plausible velocity gradients throughout the atmosphere. This comprehensive visualization confirms that our inversion approach keeps physical consistency across all parameters while handling the full range of total energies available in the RADYN dataset.

\section{Application on CHASE Spectral Observations}
\label{sec:application}

\subsection{Single Pixel Inversion}
\label{subsec:single_pixel}

\begin{figure}[ht!]
    \centering
    \includegraphics[width=\linewidth]{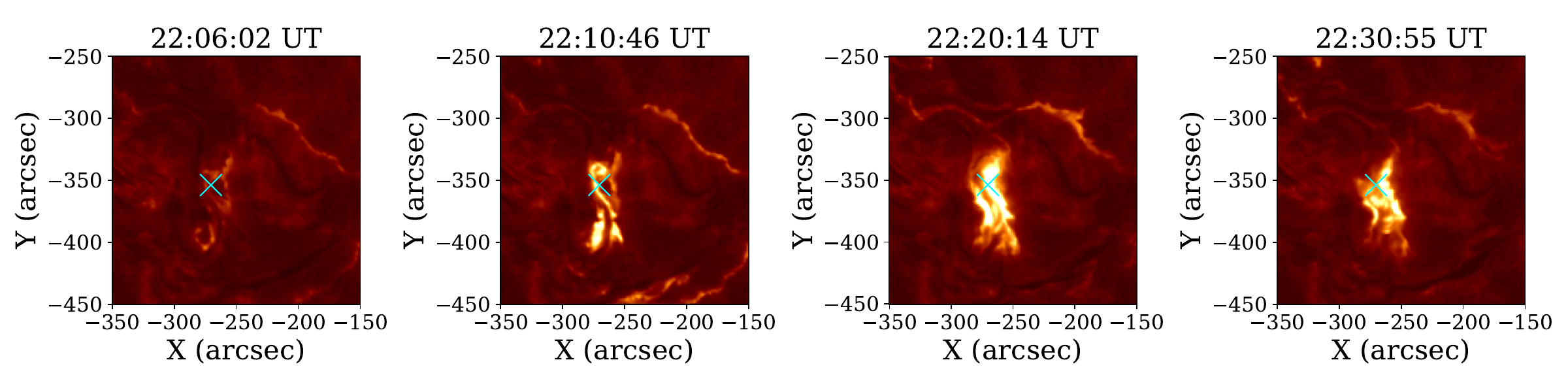}
    \caption{Images of the class X7.1 solar flare event occurring on October 1, 2024. The four panels correspond to the background, rising phase, impulsive phase, and decay phase. Cyan crosses mark the point of maximum brightness within the flare region.}
    \label{Flare}
\end{figure}

The models are then applied to real observational data, the CHASE H$\alpha$ spectra, to assess their performance in real-world scenarios. For analysis, a typical X7.1-class solar flare that occurred in active region 13842 on October 1, 2024, is selected. The flare started at approximately 21:58 UT, peaked at 22:20 UT, gradually decayed, and ended around 22:29 UT. Figure \ref{Flare} shows the evolution of the flare, including the pre-flare background, rising phase, impulsive phase, and decay phase. The pixel with the maximum brightness within the region is marked by the cyan crosses in the figures.

\begin{figure}[ht!]
    \centering
    \includegraphics[width=\linewidth]{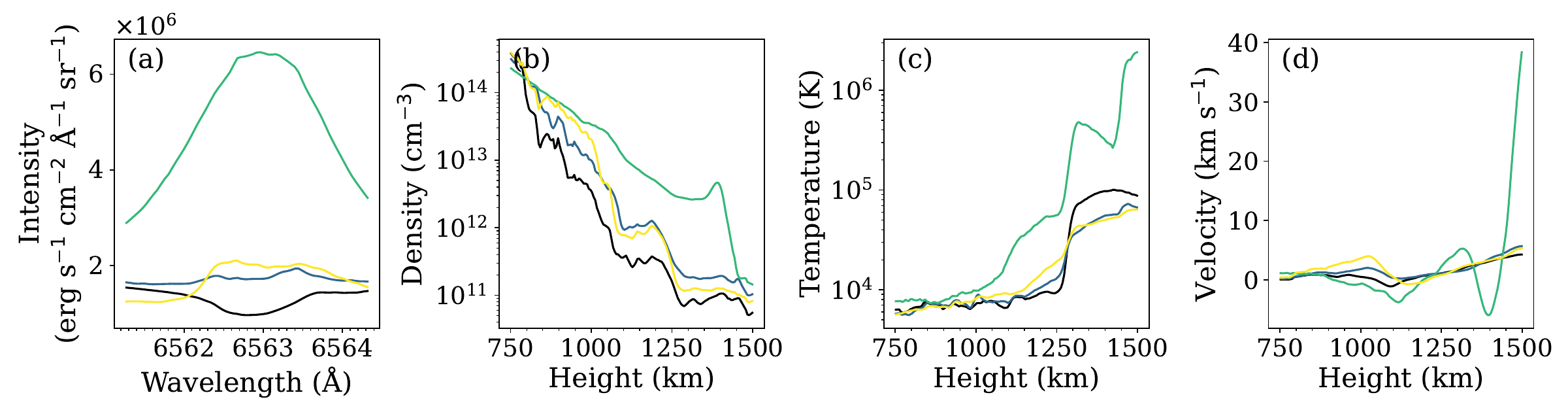}
    \caption{Inversion results for the point of maximum brightness within the region using CHASE observational data. (a) Spectra; (b) Electron number density distributions; (c) Temperature distributions; (d) Plasma velocity distributions. Different colors indicate distinct evolutionary phases: background (black), rising phase (blue), impulsive phase (green), and decay phase (yellow), respectively.}
    \label{Performance on CHASE Spectra}
\end{figure}

The spectra at the pixel in the different evolutionary phases are then inputted into the inversion models. These models then infer the corresponding distributions of physical parameters, i.e. electron number density, temperature, and plasma velocity. Considering the heights at which the H$\alpha$ spectral lines form, we display the inverted distributions at heights of 750-1500 km. The inversion results are shown in Figure \ref{Performance on CHASE Spectra}. We find that during the flare peak, the upper chromosphere is heated to the coronal temperature, and there seems to be an obvious chromospheric condensation region near 1400 km. 

\subsection{Flare Region Inversion}
\label{subsec:flare_region}

Although our model is based on one-dimensional RADYN simulation data, the spatial resolution of the CHASE spectral observations allows us to apply the model to the entire flare region without considering interactions between adjacent points. To this end, we conducted applications that can reconstruct the physical information of the flare region to a certain extent.

\begin{figure}[ht!]
    \centering
    \includegraphics[width=\linewidth]{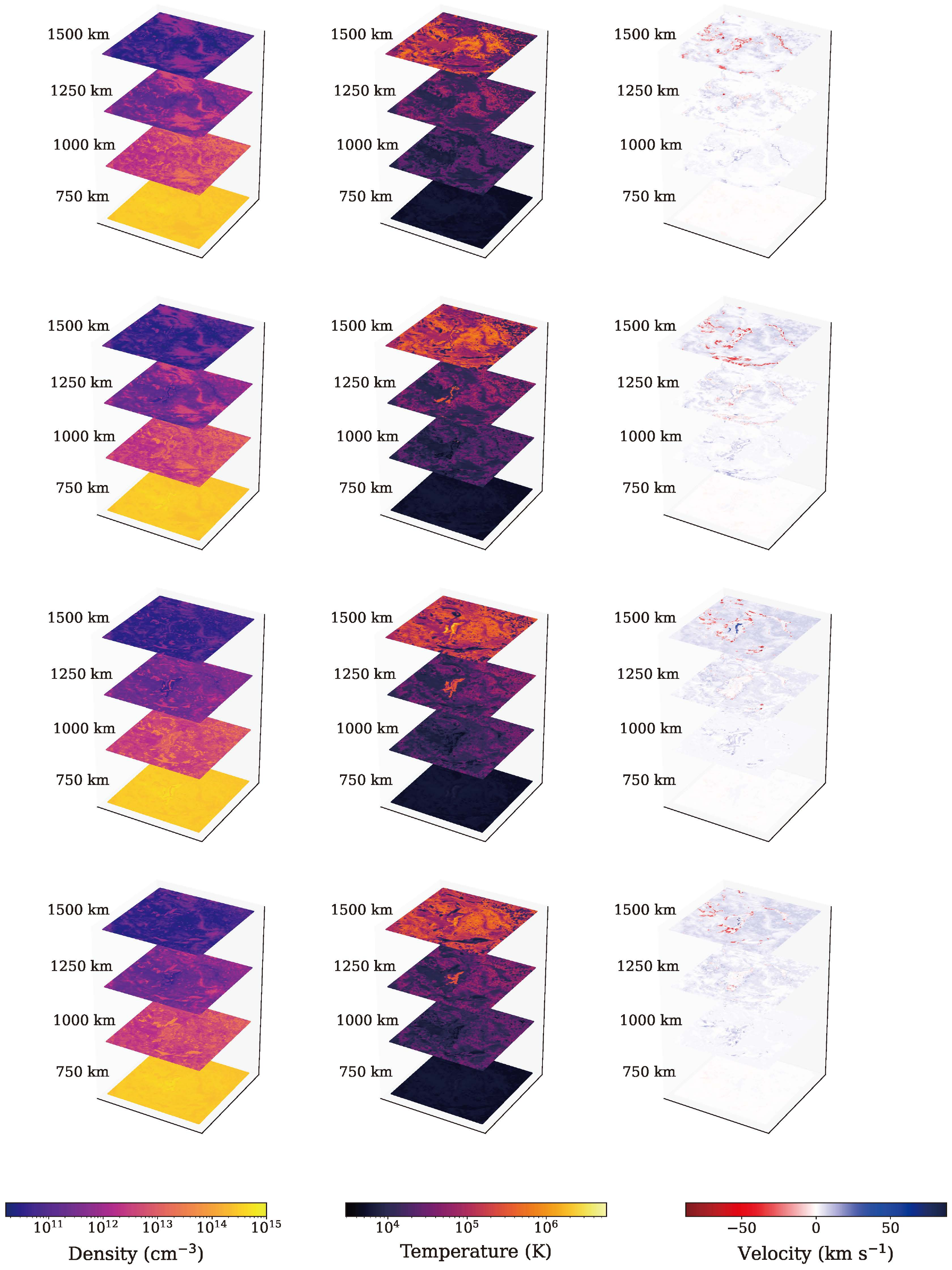}
    \caption{Height distribution characteristics of inverted physical parameters in the solar atmosphere within the flare region. The four rows from top to bottom correspond to the background, rising phase, impulsive phase, and decay phase, while the three columns from left to right represent electron number density, temperature, and plasma velocity, respectively.}
    \label{Distributions_20241001}
\end{figure}

We still use the October 1, 2024 event for our study as an example. We extract CHASE spectral data of the flare, covering the entire flare region and its different evolutionary phases. These spectra are then input into models that invert the distributions of electron number density, temperature, and plasma velocity at various heights. Four slices are selected for display at heights of 750, 1000, 1250, and 1500 km. Figure \ref{Distributions_20241001} shows the height distribution characteristics of the inverted physical parameters in the solar atmosphere within the flare region. The output parameters of our models reflect the physical changes in the different layers of the solar atmosphere during the solar flare. These 3D representations provide a visual understanding of these phenomena and aid in comprehending flare mechanisms.

\section{Summary}
\label{sec:summary}

In this study, we introduce a deep learning framework for inverting physical parameters from H$\alpha$ spectral profiles. This framework is based on simulated flare data from RADYN and observations from the CHASE satellite. Training three separate FCNNs enables us to invert electron number density, temperature, and plasma velocity from spectra with high accuracy.

These models perform well with unseen synthetic test data, accurately capturing the magnitude and structure of height-dependent profiles under various flare conditions. Of the three parameters, electron number density inversion shows the most consistent performance, while plasma velocity, though slightly noisier, still reproduces general trends. These results suggest that the models can learn nonlinear relationships between spectral features and physical quantities under NLTE conditions.

Applying the models to CHASE observations of an X-class flare demonstrates their effectiveness in real-world scenarios. The inverted parameters show consistent variation across flare phases and provide a plausible picture of the solar atmosphere response. This application highlights the potential of combining simulation-based training with deep learning to bridge the gap between physical models and observational data.

It is important to note that our method uses a FCNN whose results are significantly influenced by the training data, specifically the outputs from the RADYN simulations. If scenarios are encountered that are not represented in the RADYN simulation data (e.g., observational spectra that are not included in the synthetic dataset), the model may fail to accurately reproduce the distribution of physical parameters. Therefore, flares with anomalous spectral profiles may require more sophisticated models for analysis. One example is the INN model proposed by \citet{Osborne2019}, which incorporates additional spectral information and combines RADYN simulations with H$\alpha$ and Ca II spectra to improve parameter inference. We can improve the model by including additional spectral lines and expanding the training dataset.

Furthermore, errors were observed in the inversion results for certain time slices during the rising or decay phases of a flare. For example, discrepancies were observed between the temperature and density in the inversion results for a fixed spectral line intensity. This requires examining the observational characteristics of multiple spectral lines or incorporating additional physical constraints to eliminate such anomalies. As our model is based on RADYN flare simulations with a vertical viewing angle, it would be best applicable to flare ribbon regions near the disk center. This method is unreliable for other regions, such as flare loops, and for other events occurring over the flare ribbon, such as filament eruptions. Caution should be exercised when applying the model to flare ribbons near the solar limb, since it does not account for any center-to-limb variations of the spectral lines. Additionally, the spatial alignment of the spectral data must be carefully considered because neglecting this can result in misplaced inverted results and consequently generate inaccuracies. When interpreting the inversion results, it is important to note that there is a degeneracy between temperature and electron density which our current model has not managed to impose a physical constraint on. 

In future work, we could explore architectures that incorporate physical constraints, such as an equation of state, to better couple density and temperature thereby preventing unphysical combinations where these quantities vary independently. Such constraints would enable the model to learn more consistent thermodynamic relationships, thereby improving the robustness of the inversion results under diverse observational conditions.

\section*{acknowledgements}

CHASE mission was supported by China National Space Administration. This work was supported by NSFC under grants 12173019, 12333009, 12127901, the CNSA project D050101, the Fundamental Research Funds for the Central Universities KG202506, and the Young Data Scientist Program of the China National Astronomical Data Center, as well as the AI \& AI for Science Project of Nanjing University. J.H. is funded by the
European Union through the European Research Council
(ERC) under the Horizon Europe program (MAGHEAT, grant
agreement 101088184). The Institute for Solar Physics is
supported by a grant for research infrastructures of national
importance from the Swedish Research Council (registration
number 2021-00169).

\section*{Author Contributions}
Q. Hao conceived the ideas and supervised the project. W. Xu implemented the study and wrote the initial draft. Z. Zheng and J. Hu improved the model. J. Hong supervised the spectral inversion. Y. Qiu processed the CHASE spectral data. C. Li, M. D. Ding and C. Fang initiate the study and joined the discussions. All authors read and approved the final manuscript.
 
\section*{Declaration of Interests}
The authors declare no competing interests.

\appendix                  

\section{Inversion Results for M-class and C-class Flares}
\label{sec:appendix}

To further assess the generalization performance of our proposed method across a range of flare magnitudes, we applied the models to M-class and C-class flares. We strictly adhered to the same processing and inversion workflow utilized for the X7.1-class flare, as described in Section \ref{sec:application}.  This consistent approach ensures that the results presented here are directly comparable with those detailed in the main text,  demonstrating the robust performance of the model across various real-world scenarios.

\subsection{M-class Flare Event}

\begin{figure}[ht!]
    \centering
    \includegraphics[width=\linewidth]{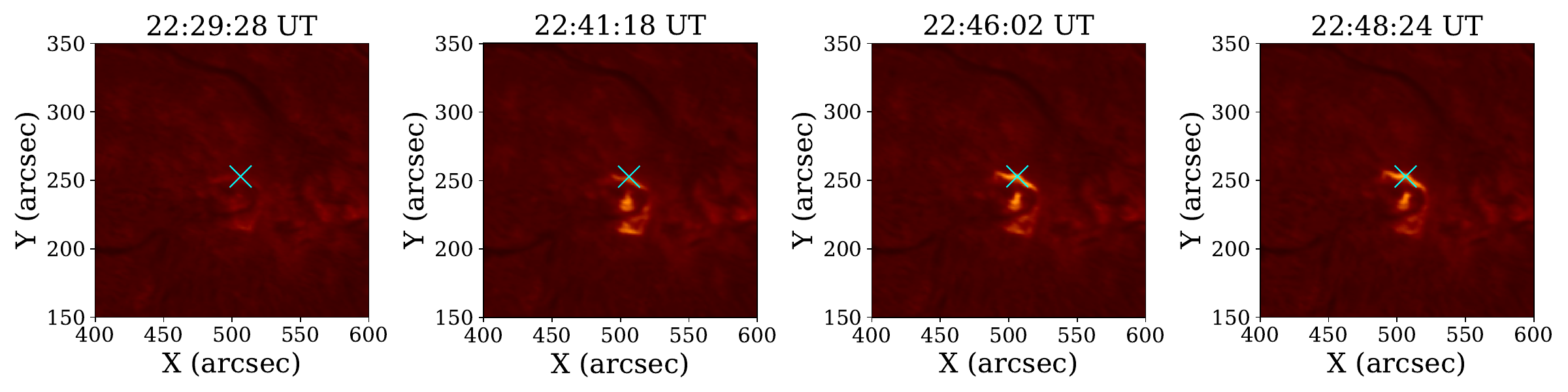} 
    \caption{Images of the class M9.0 solar flare event occurring on January 10, 2025. The four panels correspond to the background, rising phase, impulsive phase, and decay phase. Cyan crosses mark the point of maximum brightness within the flare region.}
    \label{Flare_M}
\end{figure}

For the M-class example, the M9.0-class solar flare that occurred in active region 13947 on January 10, 2025 was selected. The flare started at around 22:29 UT, peaked at 22:46 UT, and ended at approximately 22:53 UT. Figure \ref{Flare_M} shows the evolution of the flare, including the pre-flare background, rising phase, impulsive phase, and decay phase. The pixel with the maximum brightness within the region is marked by the cyan crosses in the figures.

\begin{figure}[ht!]
    \centering
    \includegraphics[width=\linewidth]{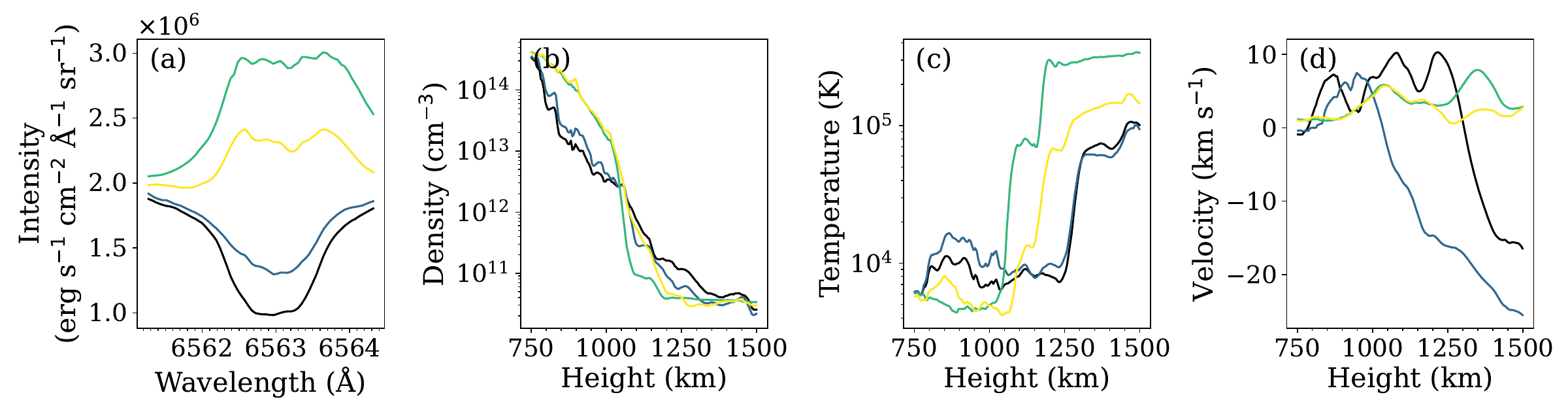}
    \caption{Inversion results for the point of maximum brightness within the region using CHASE observational data. (a) Spectra; (b) Electron number density distributions; (c) Temperature distributions; (d) Plasma velocity distributions. Different colors indicate distinct evolutionary phases: background (black), rising phase (blue), impulsive phase (green), and decay phase (yellow), respectively.}
    \label{Performance_M}
\end{figure}

Similarly to the process in Section \ref{subsec:single_pixel}, the spectra at the marked pixel in the different evolutionary phases are input into the inversion models. The models then infer the distributions of  the corresponding physical parameters, i.e. electron number density, temperature, and plasma velocity. The inversion results are shown in Figure \ref{Performance_M}. We observe that the thermodynamic evolution of the chromosphere is well-captured, showing heating patterns and velocity fields that are consistent with the impulsive nature of the flare.

\begin{figure}[ht!]
    \centering
    \includegraphics[width=\linewidth]{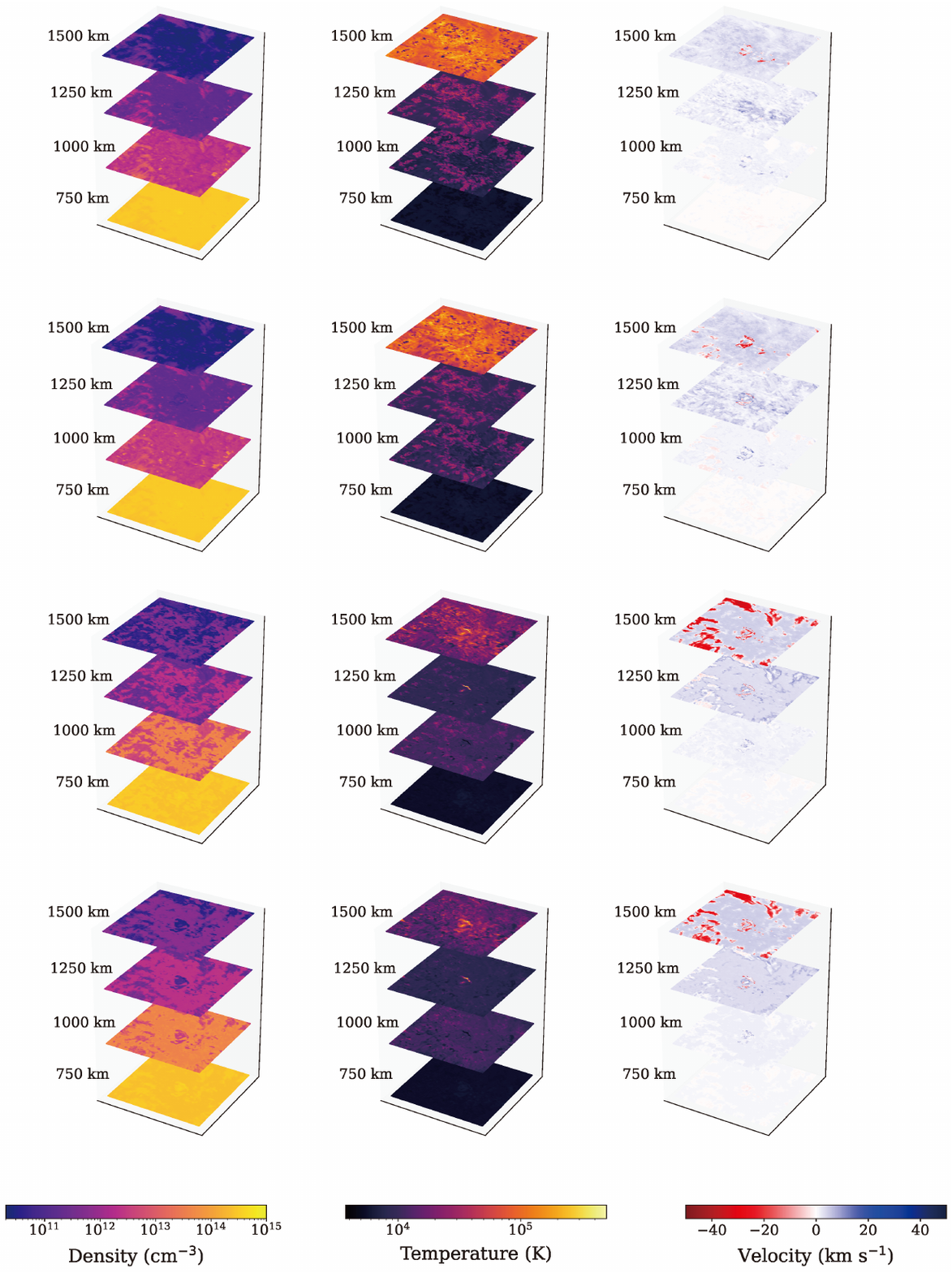}
    \caption{Height distribution characteristics of inverted physical parameters in the solar atmosphere within the flare region. The four rows from top to bottom correspond to the background, rising phase, impulsive phase, and decay phase, while the three columns from left to right represent electron number density, temperature, and plasma velocity, respectively.}
    \label{Distributions_M}
\end{figure}

Using the same strategy as in Section \ref{subsec:flare_region}, we apply the model to the entire flare region. We extracted CHASE spectral data of the flare, covering the entire flare region and its different evolutionary phases. Figure \ref{Distributions_M} shows the height distribution characteristics of the inverted physical parameters in the solar atmosphere within the flare region.

\subsection{C-class Flare Event}

\begin{figure}[ht!]
    \centering
    \includegraphics[width=\linewidth]{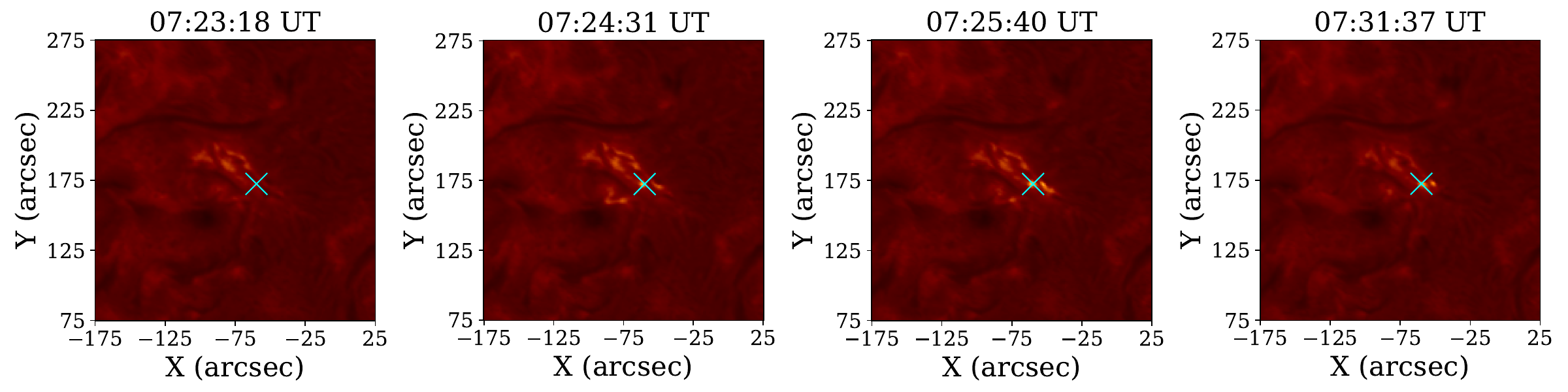}
    \caption{Images of the class C7.0 solar flare event occurring on May 31, 2025. The four panels correspond to the background, rising phase, impulsive phase, and decay phase. Cyan crosses mark the point of maximum brightness within the flare region.}
    \label{Flare_C}
\end{figure}

To evaluate performance on a weaker event, a C7.0-class solar flare that occurred in active region 14100 on May 31, 2025, was selected. The flare started at approximately 07:22 UT, peaked at 07:27 UT, and ended around 07:30 UT. Figure \ref{Flare_C} shows the evolution of the flare, including the pre-flare background, rising phase, impulsive phase, and decay phase. The pixel with the maximum brightness within the region is similarly marked in the figures.

\begin{figure}[ht!]
    \centering
    \includegraphics[width=\linewidth]{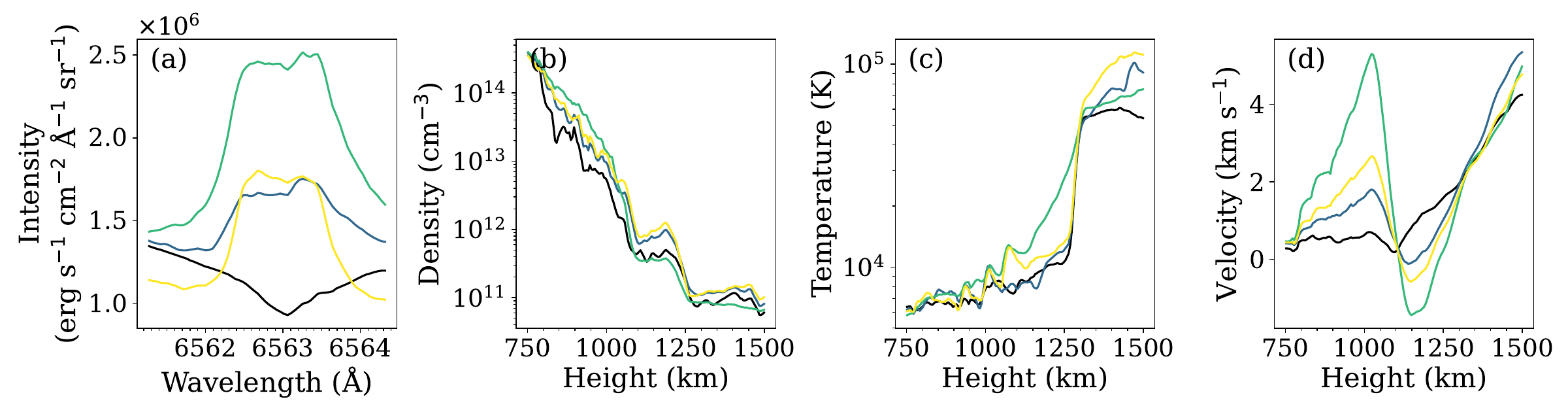}
    \caption{Inversion results for the point of maximum brightness within the region using CHASE observational data. (a) Spectra; (b) Electron number density distributions; (c) Temperature distributions; (d) Plasma velocity distributions. Different colors indicate distinct evolutionary phases: background (black), rising phase (blue), impulsive phase (green), and decay phase (yellow), respectively.}
    \label{Performance_C}
\end{figure}

The spectra from this event were processed using the same inversion pipeline. The inversion results for the single pixel are shown in Figure \ref{Performance_C}. Even for this less energetic event, the models effectively infer the distributions of physical parameters, revealing the characteristic chromospheric response to the flare energy deposition.

\begin{figure}[ht!]
    \centering
    \includegraphics[width=\linewidth]{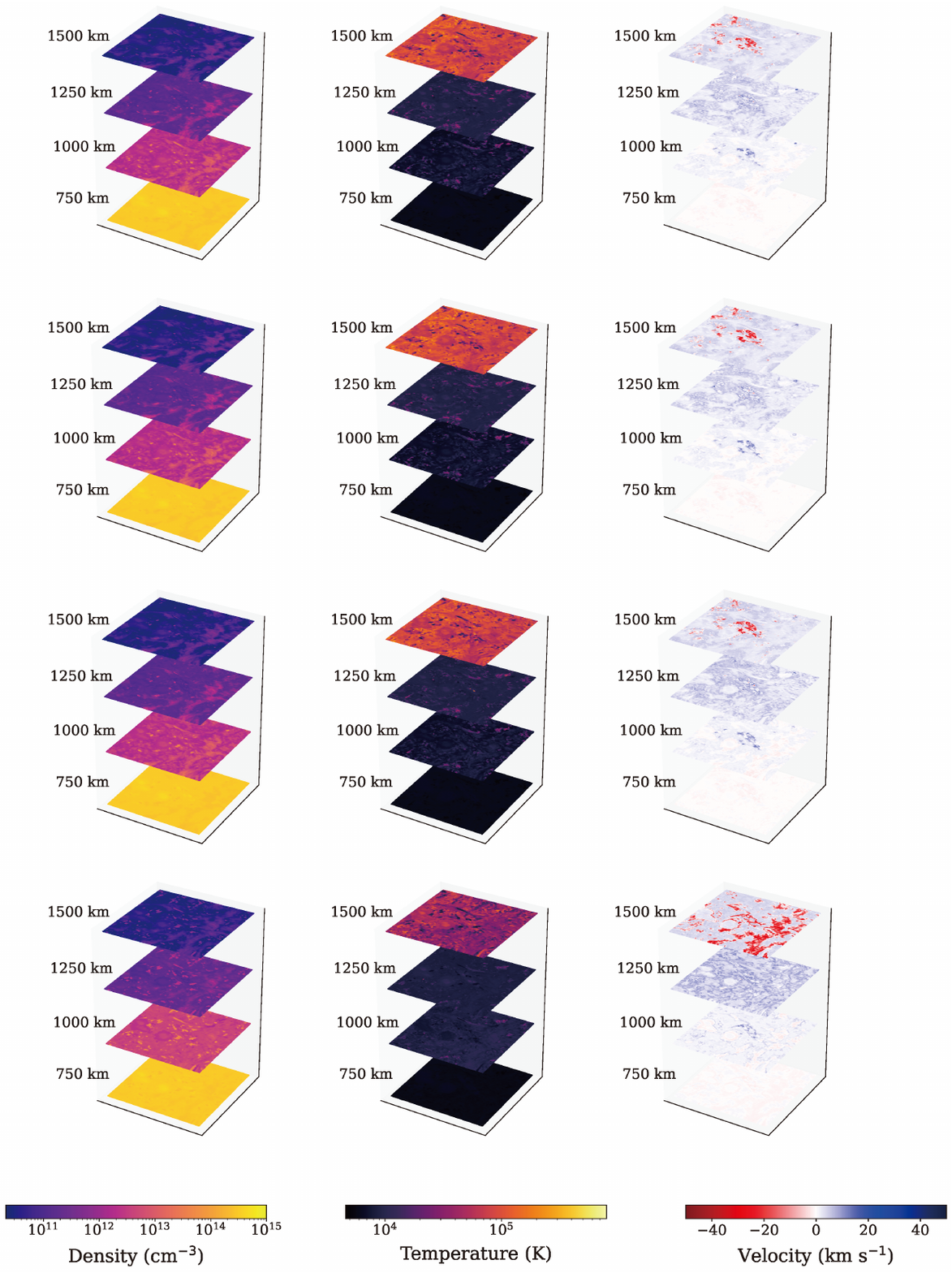}
    \caption{Height distribution characteristics of inverted physical parameters in the solar atmosphere within the flare region. The four rows from top to bottom correspond to the background, rising phase, impulsive phase, and decay phase, while the three columns from left to right represent electron number density, temperature, and plasma velocity, respectively.}
    \label{Distributions_C}
\end{figure}

Finally, the region-wide inversion is performed to reconstruct the spatial parameters. Figure \ref{Distributions_C} shows the height distribution characteristics of the inverted physical parameters. These output parameters reflect the physical changes in the different layers of the solar atmosphere during this C-class event.

\clearpage

\bibliographystyle{ati} 
\bibliography{ati}      

\label{lastpage}

\end{document}